\documentclass[prb,aps,preprint,showpacs,preprintnumbers,floats,superscriptaddress]{revtex4}
\usepackage{amssymb,amsmath,amsfonts, amsthm}
\usepackage{array,longtable}
\usepackage[mathscr]{eucal}
\usepackage[dvips]{graphicx}
\usepackage{graphics}
\newcommand{\dg}{^{\dagger}}
\begin{document}
\title{Can Frustration Preserve a Quasi-Two-Dimensional Spin Fluid?}
\date{\today}
\author{Marianna Maltseva}
\author{Piers Coleman}
\affiliation{Center for Materials Theory, Department of Physics and Astronomy,
Rutgers University, 136 Frelinghuysen Road, Piscataway, NJ 08854, USA}
\email{maltseva@physics.rutgers.edu}

\begin{abstract}

Using spin-wave  theory, we show that  geometric frustration
fails to  preserve a two-dimensional  spin fluid. Even though
frustration can remove the interlayer coupling in the ground-state of 
a classical antiferromagnet, spin layers
innevitably develop a quantum-mechanical coupling via the mechanism of
``order from disorder''. We show how the order from disorder coupling mechanism
can be viewed as  a result of magnon pair tunneling, a process 
closely analogous to
pair  tunneling in  the  Josephson  effect. In the spin system, the
Josephson coupling manifests itself as
a biquadratic spin  coupling  between layers, and for quantum spins,
these coupling terms are as large as the in-plane coupling.
An  alternative mechanism for decoupling spin layers occurs in classical
 XY models in which decoupled
"sliding  phases" of  spin fluid can form in  certain finely  tuned
conditions.  Unfortunately, 
these finely tuned  situations appear equally susceptible to
the  strong-coupling  effects  of  quantum tunneling,  forcing  us  to
conclude  that in  general,  geometric frustration  cannot preserve  a
two-dimensional spin fluid.

\end{abstract}
\maketitle
\section{Introduction}

This study is motivated by recent theories of heavy electron systems
tuned to an antiferromagnetic quantum critical point~\cite{Rosch,Si}
which propose that 
the formation of magnetically decoupled layers of spins plays
a central role in the departure from Fermi liquid behavior. 
A wide variety of heavy electron materials develop logarithmically
divergent specific heat coefficients and quasi-linear resistivities in
the vicinity of quantum critical
points~\cite{Mathur,Rosch,Aronson,Loehneysen,Julian,Custers,
Steglich,Schroeder,Rosch2,Laughlin,Belitz,Senthil,Si,Pepin,Stewart,Doiron}.
Several
theories explaining these unusual properties have been
proposed~\cite{Hertz,Moriya,Millis,Si,Pepin,
Rosch,Mathur,Belitz,Laughlin,Senthil}.  The standard model for these
quantum phase transitions, proposed by Hertz and Moriya, involves a
soft, antiferromagnetic mode coupled to a Fermi surface.  
Hertz-Moriya SDW theory can account for the 
logarithmically divergent specific heat coefficients and quasi-linear
resistivities~\cite{Rosch,Indranil}, but only if the spin fluctuations are
quasi-two-dimensional.  An alternative local quantum critical
description, based on the extended dynamical mean field theory, also
requires a quasi-two-dimensional spin fluid~\cite{Si}.  Each of these
theories can only account for the anomalies of quantum critical heavy
electron materials if the spin fluctuations of these systems are
quasi-two-dimensional~\cite{Mathur,Rosch,Hertz,Moriya,Millis,Pepin}.

\begin{figure}
\includegraphics[angle=0,width=0.7\linewidth]{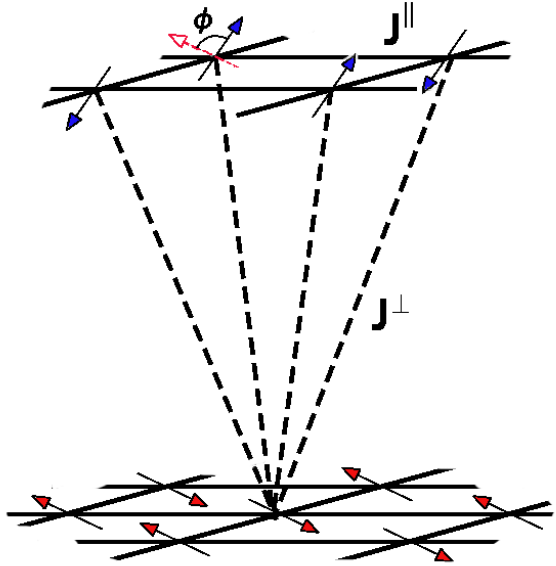}
\caption{\label{lattice} Lattice Structure.}
\end{figure}

The hypothesis that heavy electrons involve decoupled layers of spins
motivates a search for a mechanism that might preserve
quasi-two-dimensionality in a diverse set of heavy fermion materials. One such
frequently cited mechanism 
is geometric frustration~\cite{Mathur,Rosch}. Here, the
idea is that frustration, naturally induced by the structure of the
crystal, decouples layers of spins within the
material~\cite{Mathur,Rosch} (see Fig.~\ref{lattice}).
In this paper, using
the Heisenberg antiferromagnet as a simple example to explore this
line of reasoning, we show 
with the help of spin-wave theory that in
general, zero-point fluctuations of the spin overcome the frustration
and generate a strong
interlayer coupling via the mechanism of
``order from disorder'' ~\cite{Henley, Shender}. 

To illustrate the main points of our argument, consider two separate layers of Heisenberg spins.
At $T=0$ each layer is antiferromagnetically ordered, and spin waves run along the layers. Now
consider the effect of a small frustrated interlayer coupling. In a system of classical spins,
the layers remain decoupled in the classical ground state, and their spins may be rotated independently.
The long-wavelength spin waves continue to run along the layers, and the spin fluid is quasi-two-dimensional
at long wavelengths.
 
In the quantum-mechanical picture, even a small interlayer coupling enables magnons to virtually
tunnel between layers. An antiferromagnet can be regarded as a long-range RVB state~\cite{Doucot},
so individual magnon transfer is energetically unfavorable, and the transfer of magnons between
the layers tends to occur in pairs, as in Josephson tunneling (see Fig. 2).
Interlayer magnon pair tunneling
is ubiquitous in three-dimensional spin systems, frustrated and unfrustrated alike. So unless the
interlayer coupling constant is set exactly to zero, magnons travel
between the layers, producing a coupling closely analogous to
Josephson coupling of superconducting layers. Such a coupling is an
alternative way of viewing the phenomenon of ``order from disorder''
\cite{Henley, Shender}, whereby the free energy of zero-point or thermal fluctuations
depends on the relative orientation of the classical magnetization.

If we use the analogy between superconductors and antiferromagnets,
then spin rotations 
of an antiferromagnet map onto gauge transformations of the electron phase 
in a superconductor. In a superconducting tunnel junction,
the Josephson energy is determined by the product of the order
parameters in the two layers, i.e.
\[
\Delta E_J\sim - \frac{t_{\perp}^{2}}{\Delta }{\rm  Re}\left[
\langle \psi\dg _{2\uparrow}\psi\dg_{2\downarrow }\rangle 
\langle \psi _{1\uparrow}\psi\dg_{1\downarrow }\rangle 
  \right]\propto \cos(\phi_2-\phi_1)
\]
where $t_{\perp }$ is the tunneling matrix element, $\Delta $ the
superconducting gap energy and $\psi_{l \sigma }$ ($l=1,2$) represents
an electron field in lead one and two. 
By analogy, in a corresponding ``spin junction'',  the coupling energy
is determined by the product of the spin-pair
amplitudes. Suppose for simplicity that  the system is an
easy-plane XY magnet, then
\[
\Delta E_S (\Delta \phi )\sim 
- \frac{J_{\perp}^{2}}{J_{\parallel }}
{\rm  Re}\left[
\langle S^{+}_{2} (i)S^{+}_{2} (j)
\rangle 
\langle S^{-}_{1} (i')S^{-}_{1} (j')
  \right]\propto - \frac{J_{\perp}^{2}}{J_{\parallel }}
\cos(2 \Delta \phi), 
\]
where $S^{\pm}_{li}$ represents the  spin raising, or lowering
operator at site $i$ in plane $l$, parallel to the local
magnetization. 
The factor $2\Delta \phi$ arises because the 
spin-pair carries a phase which is twice the angular
displacement of the magnetization
($S_{+}\equiv S_{x}+iS_{y}\sim S e^{i\phi }$). 
In other words, 
\[
\Delta E_S (\Delta \phi )\sim 
- \frac{J_{\perp}^{2}}{2J_{\parallel }}
\cos^{2}(\Delta \phi)+\hbox{const},
\]
so the interlayer coupling induced by spin tunneling is expected
to be biquadratic in the relative angle between the spins.
Clearly, this is a much oversimplified argument. We need to take
account of the $O (3)$, rather than the $U (1)$ symmetry of a
Heisenberg system. Nevertheless, this simple argument captures the
spirit of the coupling between spin layers, as we shall now see in a
more detailed calculation. 

\begin{figure}
\includegraphics[width=0.6\linewidth]{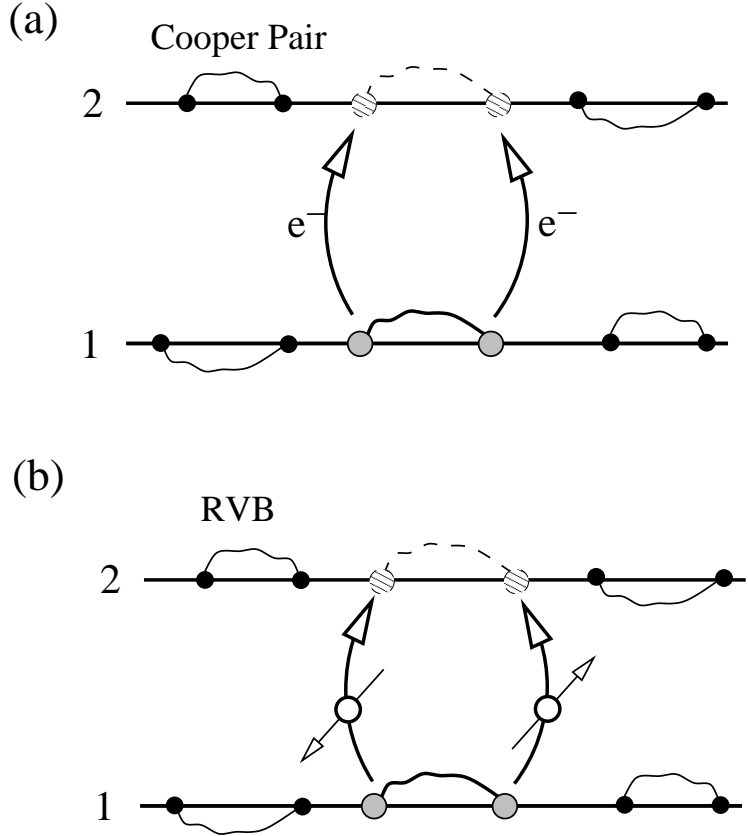}
\caption{\label{tunnel}Contrasting  (a) Josephson tunneling between paired
superconductors  and (b) magnon tunneling between antiferromagnets,
viewed within a resonating valence bond (RVB) picture.}
\end{figure}

\section{Spin-Wave Spectrum for Decoupled Layers}

Consider a Heisenberg model with nearest-neighbor antiferromagnetic interaction in its ground state defined on
the body-centered tetragonal lattice. This choice of model is motivated by the structure of $CePd_2Si_2$,
one of the compounds for which the idea of quasi-two-dimensionality was originally proposed~\cite{Mathur}.  
In this lattice structure (Fig.~\ref{lattice}), square lattices stack with a shift of ($\frac {a}{2}$,
$\frac {a}{2}$) between adjacent layers ($a$ is the lattice constant within the layer). For simplicity,
the distance between the layers is also $a$. The spins of the nearest neighbors in each layer are anti-parallel.
In the classical ground state the spins in different layers are decoupled and may assume any relative alignment.

For simplicity, let us consider just two adjacent layers, the argument being
easily generalized to an infinite
number of layers. The Hamiltonian is then

\begin{equation}
H=H_0+V,
\label{F1}
\end{equation} with

\begin{equation}
H_0=H^{(B)}+H^{(T)},
\end{equation} where $H^{(T)}$ and $H^{(B)}$
are the Hamiltonians for the top and bottom layers, and $V$ is the interlayer coupling.

\begin{equation}
H_0= J^\|\sum_{\bf i,\Delta}\Bigl({\bf S}_{\bf i}^{(B)}{\bf S}_{\bf
i+\Delta}^{(B)}+
{\bf S}_{\bf i+\delta}^{(T)}{\bf S}_{\bf i+\delta+\Delta}^{(T)}\Bigr),\\
\label{F2a}
\end{equation}

\begin{equation}
V=J^\perp\sum_{\bf i,\Delta}\Bigl({\bf S}_{\bf i}^{(B)}{\bf S}_{\bf i\pm\delta}^{(T)}+
{\bf S}_{\bf i}^{(B)}{\bf S}_{\bf i+\delta-\Delta}^{(T)}\Bigr).
\label{F2b}
\end{equation}

Here ${\bf S}_{\bf i}^{(B)}$  (${\bf S}_{\bf i}^{(T)}$) is the spin variable defined at the site ${\bf i}$
in the bottom (top) layer. The vector ${\bf \Delta}$ denotes a displacement to the nearest neighbor sites
within the plane, ${\bf \Delta}=(a,0)$ or $(0,a)$. ${\bf \delta}=(a/2,a/2)$ defines a shift between layers.
Since the coupling between layers is small ($J^\perp \ll J^\|$), we may treat this model using perturbation
theory where the ratio of coupling constants $J^{\perp}/J^{\|}$ is taken as a small parameter.
 
For our purposes, it is sufficient to consider a simple case with the spins lying in the planes of the
2-dimensional lattice. At sites ${\bf i}=(la,ma)$ and ${\bf i+\delta}=(la+\frac{1}{2}a,ma+\frac{1}{2}a)$
the spins are

\begin{eqnarray}
S_{\bf i}^{X(B)}=S(-1)^{l+m},\;\;\;
S_{\bf i}^{Y(B)}=0;\\
S_{\bf i+\delta}^{X(T)}=S (-1)^{l+m+1}\cos \phi,\;\;\;\;\;
S_{\bf i+\delta}^{Y(T)}=S (-1)^{l+m+1}\sin \phi;
\end{eqnarray} where $X$ and $Y$ are mutually perpendicular directions in the plane, $l$ and $m$ are integers.

Following a standard procedure~\cite{Cooper,Premi}, we use the Holstein-Primakoff approximation for the spin
operators to determine the spin-wave spectrum. The single-layer Hamiltonian $H^{(\alpha)}$ becomes 

\begin{eqnarray}
H^{(\alpha)} & = & -4NS^{2}J^{\|}+ \sum_{\bf q} \Bigl[8SJ^\| a_{\bf q}^{+(\alpha)}a_{\bf q}^{( \alpha)}+S
J^\|({\bf q})[a_{\bf q}^{(\alpha)}a_{-\bf q}^{( \alpha)}+h.c.]\Bigr],
\end{eqnarray} and on diagonalization the Hamiltonian $H_0$ for the decoupled layers can be written as

\begin{eqnarray} 
H_0 & = & E_0+\sum_{\alpha=T,B}\sum_{\bf q}\omega^{\|}_{\bf q} b_{\bf q}^{+(\alpha)} b_{\bf q}^{(\alpha)}.
\end{eqnarray} The ground state energy of the decoupled two-layer system is then

\begin{eqnarray} 
E_0 & = &-8NS(S+1)J^\|+\sum_{\bf q} \omega^{\|}_{\bf q}.
\end{eqnarray} $\omega^{\|}_{\bf q}$ defines the spectrum of spin waves propagating in each of the layers

\begin{equation} 
\omega_{\bf q}^{\|}=4SJ^\|\sqrt{ 4 - [\cos q_\xi a+\cos q_\eta a]^{2}}.
\end{equation} 

\section{Magnon Pair Tunneling between the Layers}

Now we express the perturbation $V$ in (\ref{F2b}) in terms of $a_{\bf q}^{+(\alpha)} $, $a_{\bf q}^{(\alpha)}$
as

\begin{eqnarray}
V =S \sum_{\bf q} A^\perp_{\bf q}(a_{\bf q}^{+(T)}a_{\bf q}^{(B)}+h.c.)+
S \sum_{\bf q} B^\perp_{\bf q}(a_{\bf q}^{(T)}a_{-\bf q}^{(B)}+h.c.),
\end{eqnarray} where $A^\perp_{\bf q}$ and $B^\perp_{\bf q}$ are defined as

\begin{equation} 
A^\perp_{\bf q}=
2J^\perp\Bigl[\cos(\frac{q_{\xi}a}{2})\cos(\frac{q_{\eta}a}{2})+\sin(\frac{q_{\xi}a}{2})\sin(\frac{q_{\eta}a}{2})\cos\phi\Bigr],
\end{equation} 

\begin{equation} 
B^\perp_{\bf q}=
2J^\perp\Bigl[\cos(\frac{q_{\xi}a}{2})\cos(\frac{q_{\eta}a}{2})-\sin(\frac{q_{\xi}a}{2})\sin(\frac{q_{\eta}a}{2})\cos\phi\Bigr].
\end{equation}  
In terms of $b_{\bf q}^{+(\alpha)} $, $b_{\bf q}^{(\alpha)}$ 

\begin{eqnarray}
V& =&S \sum_{\bf q} \alpha_q
(b_{\bf q}^{(T)}b_{-\bf q}^{(B)}+b_{\bf q}^{+(T)}b_{-\bf q}^{+(B)})+V_{ph},
\end{eqnarray} where $\alpha_q=(A^\perp_{\bf q}\sinh 2u_{\bf q}+B^\perp_{\bf q}\cosh 2u_{\bf q})$ describes the
amplitude for magnon pair tunneling, and
$V_{ph}=S \sum_{\bf q} (A^\perp_{\bf q}\cosh 2u_{\bf q}+B^\perp_{\bf q}\sinh 2u_{\bf q}) 
(b_{\bf q}^{+(T)}b_{\bf q}^{(B)}+b_{\bf q}^{+(B)}b_{\bf q}^{(T)})$
describes single magnon tunneling between layers.

It is straightforward to see that the particle-hole terms do not affect the ground-state energy, for
$V_{ph}|GS>=0$, where $|GS>$ denotes the ground state wave-function of the system. The second order correction
to the ground state energy $E_{0}$ is then

\begin{eqnarray} 
\Delta E_0^{(2)}=\sum_{\lambda}\frac{|<\lambda|\hat V|GS>|^2}{E_\lambda-E_{GS}},
\end{eqnarray} where $|\lambda>$ denotes a state with two magnons being transfered between layers. Thus,

\begin{eqnarray} 
\Delta E_0^{(2)}=-\sum_{\bf q}S^{2}\alpha_q^2/(2\omega_{\bf q}^{\|}).
\end{eqnarray}

To understand the nature of coupling between the layers (dipolar or quadrupolar), 
let us retrive the dependence of $\Delta E_0^{(2)}$ on the angle $\phi$.
 
\begin{eqnarray} 
\Delta E_0^{(2)}&=&-\frac{S}{2}\frac{(J^\perp)^2}{J^\|}[C_0+C_2\cos^2\phi]=\nonumber\\  
&=&-\frac{S}{2}\frac{(J^\perp)^2}{J^\|}[(C_0+\frac{C_2}{2})+\frac{C_2}{2}\cos 2\phi].
\label{F7}
\end{eqnarray}  
The particular form of the coefficients is

\begin{eqnarray} 
C_0=\int\limits_{-\pi}^\pi \frac{dx\; dy}{(2\pi)^2}\sqrt{1-\frac{1}{4}[\cos x+\cos y]^2}
\cos^2 \frac{x}{2}\cos^2\frac{y}{2}\bigl(1-\frac{1}{2}[\cos x+\cos y]\bigr)^2,                                      
\end{eqnarray} 

\begin{eqnarray} 
C_2=\int\limits_{-\pi}^\pi \frac{dx\; dy}{(2\pi)^2}\sqrt{1-\frac{1}{4}[\cos x+\cos y]^2}
\cos^2 \phi\sin^2\frac{x}{2}\sin^2\frac{y}{2}\bigl(1+\frac{1}{2}[\cos x+\cos y]\bigr)^2.
\end{eqnarray} 

The interlayer coupling is indeed quadrupolar in nature, as foreseen earlier. Moreover,
there is no small parameter, and for small S, when $J^\perp\sim J^\|$, this coupling is not weak.
 
\section{Discussion}

In the above calculation, we considered an ordered Heisenberg
antiferromagnet at zero temperature. In practice, 
provided the spin-spin correlation length $\xi$ is
large compared with the lattice constant $a$, $\xi\gg a$, a 
biquadratic interlayer coupling will still develop. Moreover, at finite temperatures,
thermal fluctuations will produce further interlayer
coupling. 
Both thermal and quantum interlayer coupling processes 
are manifestations of ``order from disorder''. 
The main difference between the thermal and
quantum coupling processes lies in the replacement of the magnon
occupation numbers with a Bose-Einstein distribution function, and in general
both the sign and the angular dependences of the two couplings are
expected to be the same~\cite{Henley}.
In general, geometrical frustration is an extremely fragile mechanism
for decoupling spin layers and will always be overcome by quantum and
thermal fluctuations.
Our work was motivated by heavy electron systems.
These are much more complex systems  than insulating antiferromagnets,
but if our mechanism for the formation of two-dimensional spin fluid is to be frustration, it is
difficult to see how similar interlayer coupling effects might be
avoided.   
We are led to conclude that for the
hypothesis of the reduced dimensionality of the spin fluid in heavy
fermion materials to hold, a completely different decoupling mechanism
must be at work.

In the special case of XY magnetism there is, in fact, one such alternative mechanism, related to
"sliding phases". Some heavy fermion systems, such as $YbRh_2Si_2$, are XY-like, most others, such as
$CeCu_6$, are Ising-like. It is, therefore, instructive to consider whether the sliding phase mechanism
might be generalized to Heisenberg or Ising spin systems to provide an escape from the fluctuation coupling
that we have discussed.  

The existence of a "sliding phase" in weakly coupled stacks of
two-dimensional (2D) XY models was predicted by O'Hern, Lubensky and
Toner~\cite{Lubensky}. In addition to Josephson interlayer
couplings, these authors included higher-order gradient couplings
between the layers. In the absence of Josephson couplings, these
gradient couplings preserve the decoupled nature of the spin layers,
only modifying the power-law exponents of the 2D correlation
functions, $\langle S_iS_j\rangle\sim r^{-\eta}$. As the temperature is raised,
Josephson interlayer couplings become irrelevant above a particular
"decoupling temperature" $T_d$. One can always select interlayer
gradient couplings to satisfy $T_d<T_{KT}$ and produce a stable
sliding phase in the temperature window $T_d<T<T_{KT}$.

To see this in a little more detail, consider the continuous version of the Hamiltonian of two layers of
XY models, $H=H_0+V$, where $H_0$ is a sum of independent layer Hamiltonians and $V$ is the usual
Josephson-type interlayer coupling

\begin{equation}
H_0= \frac{J^{\|}}{2}\int d^2 r[\nabla_{\perp} \phi_T({\bf r})]^2+
\frac{J^{\|}}{2}\int d^2 r[\nabla_{\perp} \phi_B({\bf r})]^2,
\label {F11}
\end{equation}

\begin{equation}
V= J^{\perp}\int d^2 r\cos[\phi_T({\bf r})-\phi_B({\bf r})].
\label {F12}
\end{equation}

At low temperature, when the interlayer coupling $J^{\perp}$ is zero, the average of the intralayer spin-spin
correlation function with respect to $H_0$ is

\begin{equation}
\langle\phi^2({\bf r})\rangle_0= \eta \log(L/b),
\end{equation} and

\begin{equation}
\langle\cos[\phi({\bf r})-\phi({\bf 0})]\rangle_0\sim (L/b)^{-\eta},
\end{equation} where $\;\eta=T/2\pi J^\|$,\;  $L$ is the sample width and $b$ is a short-distance cutoff
in the XY plane.

The average of Josephson interlayer coupling $V$ scales as \! $\langle
V\rangle_0\sim L^{2-\eta}$, \;so Josephson couplings become irrelevant
at \! $T_d=4\pi J^{\|}$.  \! At temperatures above the Kosterlitz-Thouless
transition temperature \! $T_{KT}=\pi J^\|/2$, \! thermally excited
vortices destroy the quasi-long-range order and drive the system to
disorder.  In this simple example, it happens that $T_d>T_{KT}$, \;which
does not permit a sliding phase. However, higher-order gradient
interlayer couplings between the layers, when added to this model,
suppress $T_d$ below $T_{KT}$, producing a stable sliding phase for \!
$T_d<T<T_{KT}$. 
  
So can the sliding phase concept be generalized to Heisenberg spin
systems?  A sliding phase develops in the XY model
because power-law spin correlations introduce an anomalous scaling
dimension, but unfortunately, a finite temperature Heisenberg model has
no phases with power-law correlations~\cite{Polyakov}. 
In general, biquadratic interlayer couplings will always 
remain relevant in Heisenberg models.  
In the quantum-mechanical picture, as soon as a frustrated interlayer
coupling is introduced, the order-from-disorder phenomenon~\cite{Henley, Shender} generates a coupling
\! $\lambda\sim SJ^{\perp 2}/J^\|$ \! between the layers:

\begin{equation}
H=\int \frac{\rho}{2}\sum_i (\nabla \hat{n}_i)^2+ \frac{\lambda}{2}
\sum_i (\hat{n}_i-\hat{n}_{i+1})^2,
\label {F14}
\end{equation} where  \! $\rho=S^2a^2J^\|$. \! This  coupling  gives  us  a
length  scale  $l_{0}$  determined  from \! $( l_{0})^{-2}\sim\lambda/\rho$ \!
or \! $l_{0}\sim a\sqrt{S}J^\|/J^\perp$. \! Once 
the  spin  correlation  length  \! $\xi \sim a\exp(2\pi J^\|S^2/T)$ \! within  a  layer  grows to become larger
than  $l_{0}$, \; i.e. \!
$l_{0}<
a\exp(2\pi J^\|S^2/T)
$, \; a  \!  3D-ordering  phase  transition
occurs. An  estimate  of  the  3D-ordering  transition  temperature  is  then 
$T_c\sim 2\pi J^\|S^2/\ln(\sqrt{S}J^\|/J^\perp)$. \! The  answer  is
essentially  identical  in  the  classical  picture, for  here, thermal
fluctuations generate an entropic interlayer coupling\; $\lambda\sim \hbox{max}(SJ^{\perp 2}/J^\|, ~  TS^2)$, \; so  at  high
  enough  temperatures, for large S, \;
$\lambda\sim TJ^{\perp 2}/J^{\| 2}$, \;
$l_{0}\sim SaJ^{\|~3/2}/J^\perp\sqrt{T}$. \!  A classical
estimate  of  the   3D-ordering  temperature  is\;
$T_c\sim 2\pi J^\|S^2/\ln(J^\|/J^\perp)$. 

Another interesting question is whether XY models permit sliding phases at $T=0$.
The decoupling temperature, as found by O'Hern, Lubensky and Toner~\cite{Lubensky}, is

\begin{equation}
T_d(p)=\dfrac{4\pi \rho}{f_o^{-1}-f_p^{-1}}.
\end{equation}
One sees no obvious mechanism of suppressing $T_d$ to zero. A 2D sliding phase is equivalent to a 3D finite
temperature sliding phase, so the existence of a sliding phase in the XY model at zero temperature would
mean a power-law phase in 3D XY model. Since no power-law phase exists
in 3D XY-like systems, sliding phases at $T=0$ are extremely unlikely.
In conclusion, the sliding phase scenario also fails to provide a
valid general mechanism for decoupling layers in Ising-like and Heisenberg-like systems.

Let us return momentarily to consider the implications of these
conclusions for the more complex case of heavy electron materials.  It
is clear from our discussion that simple models of frustration do not
provide a viable mechanism for decoupling spin layers. One of the
obvious distinctions between an insulating and a metallic
antiferromagnet is the presence of dissipation which acts on the spin
fluctuations.   The interlayer coupling we considered here 
relies on short-wavelength spin fluctuations, and these are the ones
that are most heavily damped in a metal.  Our exclusion of such
effects does hold open a small possibility that order-from-disorder
effects might be substantially weaker in a metallic
antiferromagnet. However, if we are to take this route, then we can
certainly no longer appeal to the analogy of the insulating antiferromagnet
while discussing a possible mechanism for decoupling spin layers. 

This research is supported by the
National Science Foundation grant NSF DMR 0312495.  We should
particularly like to thank Tom Lubensky for a discussion relating to the
sliding phases of XY antiferromagnets.

\end{document}